\documentclass[prl,superscriptaddress,twocolumn,letterpaper]{revtex4-1}
\usepackage{bm,graphicx,graphics,amsmath,amssymb,epsfig,color}
\usepackage{euscript,tabularx}
\usepackage{longtable}
\usepackage{MnSymbol,wasysym,stmaryrd,ifsym,rotating}

\begin{document}

\title{Quantitative MRFM characterization of the autonomous and forced
  dynamics\\ in a spin transfer nano-oscillator}

\author{A. Hamadeh} \affiliation{Service de Physique de l'\'Etat
  Condens\'e (CNRS URA 2464), CEA Saclay, 91191 Gif-sur-Yvette,
  France}

\author{G. de Loubens} \email[Corresponding author:
]{gregoire.deloubens@cea.fr} \affiliation{Service de Physique de
  l'\'Etat Condens\'e (CNRS URA 2464), CEA Saclay, 91191
  Gif-sur-Yvette, France}

\author{V.V. Naletov} \affiliation{Service de Physique de l'\'Etat
  Condens\'e (CNRS URA 2464), CEA Saclay, 91191 Gif-sur-Yvette,
  France} \affiliation{Physics Department, Kazan Federal University,
  Kazan 420008, Russian Federation}

\author{J. Grollier} \affiliation{Unit\'e Mixte de Physique CNRS/Thales
  and Universit\'e Paris Sud 11, RD 128, 91767 Palaiseau, France}

\author{C. Ulysse} \affiliation{Laboratoire de Photonique et de
  Nanostructures, Route de Nozay 91460 Marcoussis, France}

\author{V. Cros} \affiliation{Unit\'e Mixte de Physique CNRS/Thales
  and Universit\'e Paris Sud 11, RD 128, 91767 Palaiseau, France}

\author{O. Klein} \email[Co-author: ]{olivier.klein@cea.fr}
\affiliation{Service de Physique de l'\'Etat Condens\'e (CNRS URA
  2464), CEA Saclay, 91191 Gif-sur-Yvette, France}

\date{\today}

\begin{abstract}

  Using a magnetic resonance force microscope (MRFM), the power
  emitted by a spin transfer nano-oscillator consisting of a normally
  magnetized Py$|$Cu$|$Py circular nanopillar is measured both in the
  autonomous and forced regimes. From the power behavior in the
  subcritical region of the autonomous dynamics, one obtains a
  quantitative measurement of the threshold current and of the noise
  level. Their field dependence directly yields both the spin torque
  efficiency acting on the thin layer and the nature of the mode which
  first auto-oscillates: the lowest energy, spatially most uniform
  spin-wave mode. From the MRFM behavior in the forced dynamics, it is
  then demonstrated that in order to phase-lock this auto-oscillating
  mode, the external source must have the same spatial symmetry as the
  mode profile, \textit{i.e.}, a uniform microwave field must be used
  rather than a microwave current flowing through the nanopillar.
 
\end{abstract}

\maketitle

%\section{Introduction}

Recent progress in spin electronics have demonstrated that owing to
the spin transfer torque (STT) \cite{slonczewski96, berger96}, biasing
magnetic hybrid nanostructures by a direct current can lead to
microwave emission. These spin transfer nano-oscillators (STNOs)
\cite{kiselev03, rippard04, houssameddine07} offer decisive advantages
compared to existing technology in tunability, agility, compactness
and integrability. In view of their applications in high-frequency
technologies, a promising strategy to improve the coherence and
increase the emitted microwave power of these devices is to mutually
synchronize several of them \cite{kaka05, mancoff05, slavin05a,
  grollier06, ruotolo09}.

The synchronization of the STNO oscillations to an external source has
already been demonstrated \cite{rippard05,georges08}. In particular,
it has been shown that symmetric perturbations to the STNO trajectory
favor even synchronization indices (ratio of the external frequency to
the STNO frequency $r=2,4,6...$), while antisymmetric perturbations
favor odd synchronization indices \cite{quinsat11,urazhdin10}. But so
far, the influence of the spatial symmetry of the spin-wave (SW) mode
which auto-oscillates on the synchronization rules has not been
elucidated.

To address this open question, the spectroscopic identification of the
auto-oscillating mode is crucial. It is usually a challenge, as a
large variety of dynamic modes can be excited in STNOs, and their
nature can change depending on the geometry, magnetic parameters and
bias conditions. In this work, we study a STNO in the most simple
configuration: a circular nanopillar saturated by a strong magnetic
field applied along its normal. It corresponds to an optimum
configuration for synchronization, since it has a maximal nonlinear
frequency shift, which provides a large ability for the STNO to lock
its phase to an external source \cite{slavin05a}. Moreover, the
perpendicular configuration coincides with the universal oscillator
model, for which an exact analytical theory can be derived
\cite{slavin09}. Last but not least, this highly symmetric case allows
for a simplified classification of the SW eigenmodes inside the STNO
\cite{naletov11}.

We shall use here a magnetic resonance force microscope (MRFM) to
monitor directly the power emitted by this archetype STNO vs. the bias
dc current and perpendicular magnetic field. In the autonomous regime,
these quantitative measurements allow us to demonstrate that the mode
which auto-oscillates just above the threshold current is the
fundamental, spatially most uniform SW mode. By studying the forced
regime, we then show that this mode synchronizes only to an external
source sharing the same spatial symmetry, namely, a uniform microwave
magnetic field, and \emph{not} the common microwave current passing
through the device.

%\section{Experimental setup}

For this study, we use a circular nanopillar of nominal diameter
200~nm patterned from a (Cu60$|$Py$_B$15$|$ Cu10$|$Py$_A$4$|$Au25)
stack, where thicknesses are in nm and Py=Ni$_{80}$Fe$_{20}$. A dc
current $I_\text{dc}$ and a microwave current $i_\text{rf}$ can be
injected through the STNO using the bottom Cu and top Au electrodes. A
positive current corresponds to electrons flowing from the thick
Py$_B$ to the thin Py$_A$ layer. This STNO device is insulated and an
external antenna is patterned on top to generate a spatially uniform
microwave magnetic field $h_\text{rf}$ oriented in the plane of the
magnetic layers. The bias magnetic field $H_\text{ext}$, ranging
between 8.5 and 11~kOe, is applied at $\theta_H=0^\circ$ from the
normal to the sample plane.

The room temperature MRFM setup \cite{klein08} consists of a spherical
magnetic probe attached at the end of a very soft cantilever, coupled
dipolarly to the buried nanopillar (see inset of Fig.1) and positioned
$1.5~\mu$m above its center. This mechanical detection scheme
\cite{loubens07, pigeau11} sensitively measures the variation of the
longitudinal magnetization $\Delta M_z$ over the whole volume of the
magnetic body \cite{loubens05}, a quantity \emph{directly}
proportional to the normalized power $p$ emitted by the STNO
\cite{slavin09}:
\begin{equation}
  p=\frac{\Delta M_z}{2 M_s} \label{mrfm} \, ,
\end{equation}
where $M_s$ is the saturation magnetization of the precessing
layer.

%\section{Phase diagram of autonomous dynamics}

\begin{figure}
  \includegraphics[width=\columnwidth]{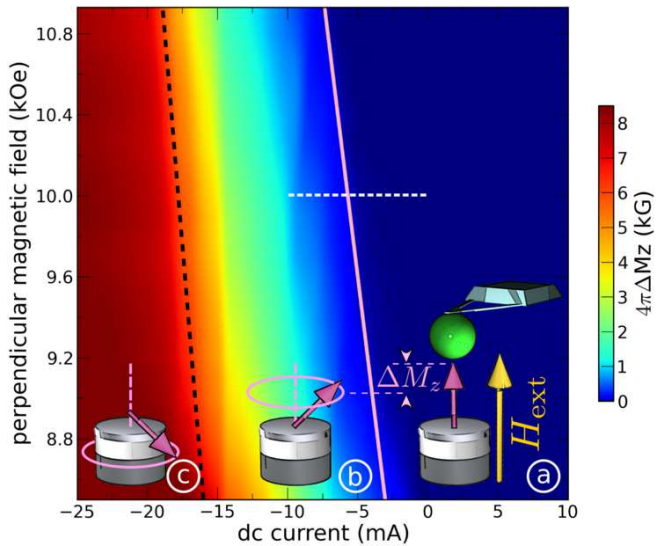}
  \caption{(Color online). Phase diagram of the STNO autonomous
    dynamics measured by MRFM.}
  \label{fig:1}
\end{figure}

First, we measure the phase diagram of the STNO autonomous dynamics as
a function of $I_\text{dc}$ and $H_\text{ext}$, see
Fig.\ref{fig:1}. In this experiment, $I_\text{dc}$ is fully modulated
at the cantilever frequency, $f_c \approx 12$~kHz, and the mechanical
signal represents $\Delta M_z$ synchronous with the injection of
$I_\text{dc}$ through the STNO. This quantitative measurement
\cite{naletov03} is displayed using the color scale indicated on the
right of Fig.\ref{fig:1}.

Three different regions can be distinguished in this phase diagram. At
low negative or positive current (region \textcircled{a}), $\Delta
M_z$ is negligible, because in the subcritical region, the STT is not
sufficient to destabilize the magnetization in the thin or thick layer
away from the perpendicular applied field direction. As $I_\text{dc}$
is reaching a threshold negative value (from $-3$ to $-7$~mA as
$H_\text{ext}$ increases from $8.5$ to $10.7$~kOe, see pink solid line
in Fig.\ref{fig:1}), the MRFM signal starts to smoothly increase in
region \textcircled{b}. It corresponds to the onset of spin transfer
driven oscillations in the thin layer, which will be analyzed in
details below. As $I_\text{dc}$ is further decreased towards more
negative values, the angle of precession increases in the thin layer,
until it eventually reaches 90\textdegree: at the boundary between
regions \textcircled{b} and \textcircled{c} (see black dashed line)
$4\pi\Delta M_z$ equals the full saturation magnetization in the thin
layer, $4\pi M_s=8$~kG.

%\section{Quantitative analysis of the subcritical regime}

\begin{figure*}
  \includegraphics[width=17cm]{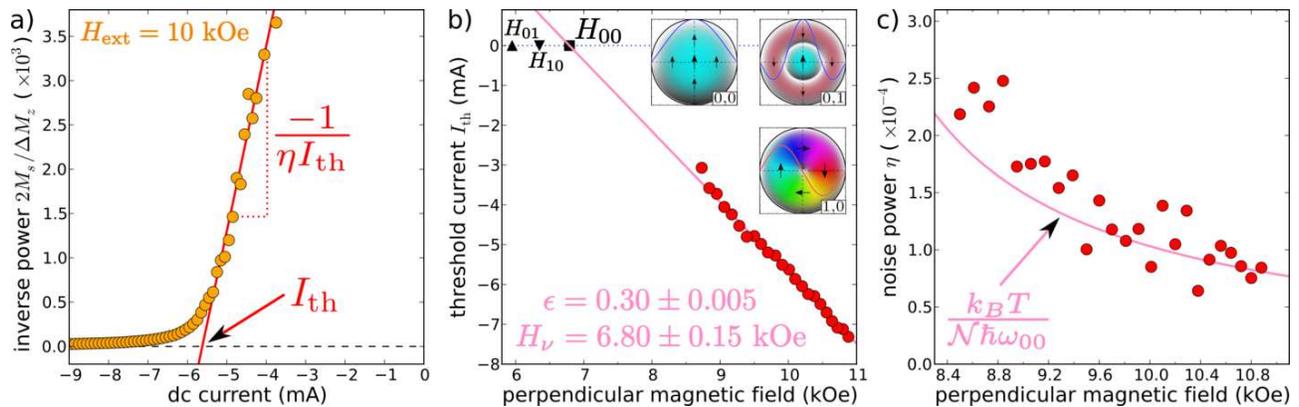}
  \caption{(Color online). (a) Determination of the threshold current
    $I_\text{th}$ and noise power $\eta$ at $H_\text{ext}=10$~kOe,
    from the inverse MRFM signal in the subcritical
    regime. Dependencies of the threshold current (b) and noise power
    (c) on the perpendicular magnetic field.}
  \label{fig:2}
\end{figure*}

Let us now concentrate on the spin transfer dynamics in the thin layer
at $I_\text{dc}<0$. We first turn to the quantitative analysis of the
subcritical region \textcircled{a}. We introduce ${\cal N}=VM_s/(g
\mu_B)$, the number of spins in the thin layer ($V$ is its volume, $g$
the Land\'e factor, $\mu_B$ the Bohr magneton). The averaged
normalized power $p$ in the subcritical regime
($|I_\text{dc}|<I_\text{th}$) is evaluated in the stochastic nonlinear
oscillator model described in section VII of
ref. \cite{slavin09}. Under the assumption that only one SW mode
dominates the STNO autonomous dynamics, Eq.(\ref{mrfm}) follows the
simple relationship:
\begin{equation}
  \frac{\Delta M_z}{2 M_s} = \frac{k_BT}{{\cal N} \hbar \omega_\nu} \, \,\frac{1}{1-I_\text{dc}/I_\text{th}}\, ,
\label{subcri}
\end{equation}
where $I_\text{th}= 2 \alpha \omega_\nu {\cal N}e/\epsilon$ is the
threshold current for auto-oscillation of the SW mode $\nu$ with
frequency $\omega_\nu$ ($\alpha$ is the Gilbert damping constant in
the thin layer, $e$ the electron charge, and $\epsilon$ the spin
torque efficiency). In Eq.(\ref{subcri}), the prefactor
\begin{equation}
  \eta \equiv \frac{k_BT}{{\cal N} \hbar \omega_\nu}
\label{eta}
\end{equation}
is the noise power: the ratio between the thermal energy ($k_B$ is the
Boltzmann constant, $T$ the temperature) and the maximal energy stored
in the SW mode $\nu$ ($\hbar$ is the Planck constant over $2\pi$).

From Eq.(\ref{subcri}), the inverse power is linear with the bias
current $I_\text{dc}$ in the subcritical region. A sample measurement
at $H_\text{ext} = 10$~kOe (along the white dashed line in
Fig.\ref{fig:1}) is shown in Fig.\ref{fig:2}a. From a linear fit, one
can thus obtain the threshold current $I_\text{th}$ and the noise
power $\eta$ at this particular field. The dependencies of
$I_\text{th}$ and $\eta$ on the perpendicular magnetic field are
plotted in Figs.\ref{fig:2}b and \ref{fig:2}c, respectively.

The parameters $V$, $M_s$, $g$ (hence, ${\cal N}\simeq6.3\times10^6$)
and $\alpha=0.014$ of the thin layer have been determined from an
extensive MRFM spectroscopic study performed at $I_\text{dc}=0$ on the
same sample and published in ref.\cite{naletov11}. This study also
yields the dispersion relations
$\omega_{\nu}=\gamma(H_\text{ext}-H_{\nu})$ of the thin layer SW modes
($\gamma=g\mu_B/\hbar=1.87\times10^7$~rad.s$^{-1}$.G$^{-1}$ is the
gyromagnetic ratio, $H_{\nu}$ the so-called Kittel field associated to
the mode $\nu$). By injecting $\omega_{\nu}$ in the expression of the
threshold current, it is found that the latter depends linearly on the
perpendicular bias field:
\begin{equation}
  I_\text{th}=\frac{2\alpha {\cal N}e}{\epsilon}\gamma(H_\text{ext}-H_\nu)\, ,
  \label{Ith}
\end{equation}
as observed in Fig.\ref{fig:2}b. The linear fit of $I_\text{th}$
vs. $H_\text{ext}$ using Eq.(\ref{Ith}) yields
$H_{\nu}=6.80\pm0.15$~kOe and $\epsilon=0.30\pm0.005$.  The importance
of the analysis of Fig.\ref{fig:2}b is that, first, it provides an
accurate determination of the spin torque efficiency, found to be in
agreement with the accepted value in similar STNO stacks
\cite{rychkov09}. Second, a comparison with the SW modes of the thin
layer (see black symbols extracted from ref.\cite{naletov11} and mode
profiles in Fig.\ref{fig:2}b) shows that the fitted value of $H_{\nu}$
precisely corresponds to the Kittel field of the $(\ell,n)=(0,0)$
mode, $\ell$ and $n$ being respectively the azimuthal and radial mode
indices. It thus allows us to conclude about the nature of the mode
that first auto-oscillates at $I_\text{dc}<0$ as being the
fundamental, most uniform precession mode of the thin layer.

To gain further insight in our analysis of the subcritical regime, we
compare in Fig.\ref{fig:2}c the noise power determined as a function
of $H_\text{ext}$ with the prediction of Eq.(\ref{eta}), in which the
dispersion relation of the $\nu=(0,0)$ SW mode is used. It is found
that the fluctuations of the STNO power are well accounted for by
those of the previously identified auto-oscillating mode, which
confirms that the single mode assumption made to derive
Eq.(\ref{subcri}) is a good approximation.

%\section{Synchronization of the auto-oscillating mode}

\begin{figure}
  \includegraphics[width=\columnwidth]{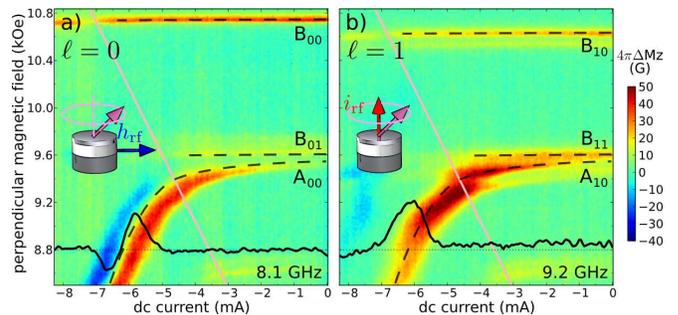}
  \caption{(Color online). MRFM measurement of the STNO dynamics
    forced by (a) the uniform field $h_\text{rf}$ at 8.1~GHz and (b)
    the orthoradial Oersted field produced by $i_\text{rf}$ at
    9.2~GHz, as a function of $I_\text{dc}$ and $H_\text{ext}$.  The
    black traces show the MRFM signal vs. $I_\text{dc}$ at
    $H_\text{ext}=8.8$~kOe. The pink solid lines show the location of
    the threshold current determined in Fig.\ref{fig:2}b. The dashed
    lines are guides to the eye.}
  \label{fig:3}
\end{figure}

Using two different microwave circuits, we shall now compare the
ability of the auto-oscillating SW mode to phase-lock either to the
uniform microwave field $h_\text{rf}$ generated by the external
antenna, or to the microwave current $i_\text{rf}$ flowing through the
nanopillar. We know from previous studies that in the exact
perpendicular configuration, the SW spectrum critically depends on the
method of excitation \cite{naletov11}: $h_\text{rf}$ excites only the
axially symmetric modes having azimuthal index $\ell=0$, whereas due
to the orthoradial symmetry of the induced microwave Oersted field,
$i_\text{rf}$ excites only the modes having azimuthal index
$\ell=+1$. The dependencies on $I_\text{dc}$ and $H_\text{ext}$ of the
STNO dynamics forced respectively by $h_\text{rf}$ and $i_\text{rf}$
are presented in Figs.\ref{fig:3}a and \ref{fig:3}b. The plotted
quantity is $\Delta M_z$ synchronous with the full modulation of the
external source power: $h_\text{rf}=1.9$~Oe (a) and
$i_\text{rf}=140$~$\mu$A (b). Although the $\ell=0$ and $\ell=+1$
spectra are in principle shifted by 1.1~GHz from each other, a direct
comparison of the phase diagrams (a) and (b) can be made by using
different excitation frequencies for $h_\text{rf}$ (8.1~GHz) and
$i_\text{rf}$ (9.2~GHz).

Below the threshold current (indicated by the pink lines in
Fig.\ref{fig:3}), the observed behaviors of the $\ell=0$ and $\ell=+1$
modes are alike: a small negative dc current slightly attenuates the
SW modes $B_{\ell n}$ of the thick Py$_B$ layer, while it promotes
quite rapidly the SW modes $A_{\ell n}$ of the thin Py$_A$ layer, in
agreement with the expected symmetry of the STT \cite{naletov11}. On
the contrary, there is a clear qualitative difference between the
modes $A_{00}$ and $A_{10}$ beyond $I_\text{th}$. Although both peaks
similarly shift towards lower field as $I_\text{dc}$ is decreased
towards lower negative values, $A_{00}$ gets strongly distorted, with
the appearance of a negative dip on its high field side, in contrast
to $A_{10}$, which remains a positive peak.

\begin{figure}
  \includegraphics[width=\columnwidth]{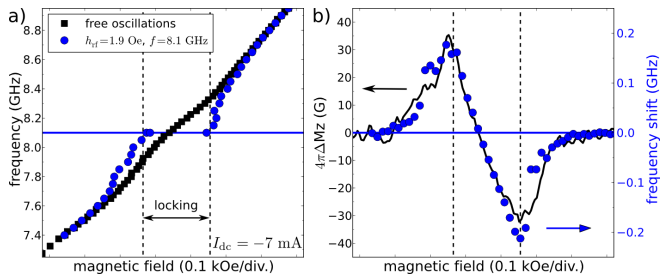}
  \caption{(Color online). (a) Magnetic field dependence of the STNO
    frequency in the free and forced regimes (the external source at
    8.1~GHz is $h_\text{rf}$). (b) Comparison between the STNO
    frequency shift deduced from (a) and the MRFM signal.}
  \label{fig:4}
\end{figure}

The negative MRFM signal observed in Fig.\ref{fig:3}a in the region of
spin transfer driven oscillations in the thin layer is striking,
because it means that the precession angle can be \emph{reduced} in
the presence of the microwave excitation $h_\text{rf}$. As a matter of
fact, the distortion of the peak $A_{00}$ is associated to the
synchronization of the auto-oscillating mode to the external
signal. Fig.\ref{fig:4}a illustrates the distortion of the STNO
emission frequency induced by this phenomenon. These data were
obtained by monitoring the fluctuating voltage across the nanopillar
at $I_\text{dc}=-7$~mA with a spectrum analyzer as a function of the
applied magnetic field \cite{Note1}. The frequency shift
of the forced oscillations with respect to the free running
oscillations is plotted in Fig.\ref{fig:4}b, along with the MRFM
signal. This demonstrates that in the so-called phase-locking range,
the STNO amplitude adapts ($\Delta M_z>0$: increases, $\Delta M_z<0$:
decreases), so as to maintain its frequency equal to the frequency of
the source, here fixed at 8.1~GHz. This comparison also allows to
estimate the phase-locking bandwidth, found to be as large as 0.4~GHz
despite the small amplitude of the external signal. The nonlinear
frequency shift is indeed the largest in the perpendicular
configuration, $N=4\gamma M_s\simeq 48$~GHz \cite{slavin09},
therefore, a small change of the power emitted by the STNO is
sufficient to change its frequency by a substantial amount.

Such a signature of synchronization of the auto-oscillating mode is
not observed in Fig.\ref{fig:3}b, where the external source is the
microwave current. This highlights the crucial importance of the
symmetry associated to the SW mode driven by STT: in the exact
perpendicular configuration, $i_\text{rf}$ can only excite $\ell=+1$
SW modes, therefore, it has the wrong symmetry to couple to the
auto-oscillating mode, which was shown in Fig.\ref{fig:2} to bare the
azimuthal index $\ell=0$.  We add that in our exact axially
symmetrical case, no phase-locking behavior is observed with the even
synchronization index $r=2$, neither with $i_\text{rf}$, nor with
$h_\text{rf}$, which is due to the perfectly circular STNO trajectory.

%\section{Conclusion}

To conclude, based on the quantitative analysis of both the critical
current and the noise power in the subcritical regime, we have
unambiguously identified the auto-oscillating mode in the
perpendicular configuration of a nanopillar. This case is particularly
interesting due to its large ability to synchronize to an external
source. But we have shown that in addition to the symmetry of the
perturbation with respect to the STNO trajectory \cite{urazhdin10},
the overlap integral between the external source and the
auto-oscillating mode profile is crucial to synchronization rules.
Due to symmetry reasons, only the uniform microwave field applied
perpendicularly to the bias field and with the synchronization index
$r=1$ is efficient to phase-lock the STNO dynamics in the present
work. We believe that this finding might be important for future
strategies to synchronize large STNOs arrays.

% --------------------------------------------------------------------
% Acknowledgments
% --------------------------------------------------------------------

We thank A.~N.~Slavin for useful discussions and his support. This
research was supported by the European Grant Master (NMP-FP7 212257)
and by the French Grant Voice (ANR-09-NANO-006-01).

%--------------------------------------------------------------------
% References
%--------------------------------------------------------------------

%

\end{document}